\documentclass{emulateapj}
\shorttitle{The Impostor SN~2008S}
\shortauthors{Smith et al.}
\begin{document}

\title{SN~2008S: A Cool Super-Eddington Wind in a Supernova Impostor}
\author{Nathan Smith\altaffilmark{1}, Mohan
  Ganeshalingam\altaffilmark{1}, Ryan Chornock\altaffilmark{1}, Alexei
  V.\ Filippenko\altaffilmark{1}, Weidong Li\altaffilmark{1}, Jeffrey
  M.\ Silverman\altaffilmark{1}, Thea N.\ Steele\altaffilmark{1},
  Christopher V.\ Griffith\altaffilmark{1}, Niels
  Joubert\altaffilmark{1}, Nicholas Y.\ Lee\altaffilmark{1}, Thomas B.\
  Lowe\altaffilmark{2}, Martin P.\ Mobberley\altaffilmark{3}, \&
  Dustin M.\ Winslow\altaffilmark{1}}

\altaffiltext{1}{Department of Astronomy, University of California,
  Berkeley, CA 94720-3411; nathans@astro.berkeley.edu .}
\altaffiltext{2}{Lick Observatory, PO Box 63, Mount Hamilton, CA
  95140.}
\altaffiltext{3}{British Astronomical Association, Burlington House,
  Picadilly, London W1J 0DU, UK.}

\begin{abstract}

  We present visual-wavelength photometry and spectroscopy of
  supernova (SN)~2008S.  Based on the low peak luminosity for a SN of
  $M_R = -13.9$ mag, photometric and spectral evolution unlike that of
  low-luminosity SNe, a late-time decline rate slower than $^{56}$Co
  decay, and slow outflow speeds of 600--1000 km s$^{-1}$, we conclude
  that SN~2008S is not a true core-collapse SN and is probably not an
  electron-capture SN.  Instead, we show that SN~2008S more closely
  resembles a ``SN impostor'' event like SN~1997bs, analogous to the
  giant eruptions of luminous blue variables (LBVs).  Its total
  radiated energy was $\sim$10$^{47.8}$ ergs, and it may have ejected
  0.05--0.2~$M_{\odot}$ in the event.  We discover an uncanny
  similarity between the spectrum of SN~2008S and that of the Galactic
  hypergiant IRC+10420, which is dominated by narrow H$\alpha$,
  [Ca~{\sc ii}], and Ca~{\sc ii} emission lines formed in an opaque
  wind.  We propose a scenario where the vastly super-Eddington
  ($\Gamma \approx 40$) wind of SN~2008S partly fails because of
  reduced opacity due to recombination, as suggested for IRC+10420.
  The range of initial masses susceptible to eruptive LBV-like mass
  loss was known to extend down to 20--25 M$_{\odot}$, but estimates
  for the progenitor of SN~2008S (and the similar NGC~300 transient)
  may extend this range to $\la$15~M$_{\odot}$. As such, SN~2008S may
  have implications for the progenitor of SN~1987A.

\end{abstract}


\section{INTRODUCTION}

The class of Type IIn supernovae (SNe~IIn) is surprisingly diverse
compared to other spectral types. SNe~IIn are classified as such
because of the relatively narrow H emission lines in their spectra
(Schlegel 1990; Filippenko 1997), but the underlying physics of the
outbursts may be quite varied.  Recent examples of extremely luminous
SNe~IIn such as SNe~2006tf and 2006gy (Smith et al.\ 2008, 2007; Ofek
et al.\ 2007) challenge our understanding of massive star evolution.

We also know of remarkably {\it faint} SNe~IIn.  It is unclear if
these belong to a tail of the core-collapse SN distribution (e.g.,
Pastorello et al.\ 2004), or if they mark a different kind of
outburst.  Among low-luminosity SNe~IIn is an observed class of
objects referred to variously as ``SN impostors'' (Van Dyk et al.\
2000), Type~V SNe (Zwicky 1965), $\eta$~Car analogs (Goodrich et al.\
1989; Filippenko et al.\ 1995), or giant eruptions of luminous blue
variables (LBVs).  These are nonterminal outbursts related to
historical eruptions of $\eta$~Car, P~Cyg, SN~1961V, and SN~1954J (see
Humphreys, Davidson, \& Smith 1999).  The physical cause of the
outbursts is unknown, but since the stars are observed to survive in
some cases (Van Dyk et al.\ 2002; Van Dyk et al.\ 2005; Smith et al.\
2001), they are thought to be distinct from core-collapse SNe.  For
distant objects, however, the case is not always clearly proven (e.g.,
Stockdale et al.\ 2001; Chu et al.\ 2004).  Conversely, Woosley et
al.\ (2007) proposed that even the most luminous SNe~IIn may be
non-terminal events.

The underlying trigger of these outbursts remains unexplained, but the
observed outflow is generally thought to be caused by violating the
classical Eddington luminosity and thereby initiating severe mass loss
(Owocki, Gayley, \& Shaviv 2004; Smith \& Owocki 2006).  These
``impostors'' exceed their pre-outburst states by several magnitudes,
with typical peak absolute visual magnitudes of $-$11 to $-$14. The
class is heterogeneous, but a representative example of a SN impostor
is SN~1997bs (Van Dyk et al.\ 2000), which was the first ``SN''
detected by the Lick Observatory SN Search (Filippenko et al.\ 2001).
Van Dyk (2005) discussed additional examples.

To this already diverse subclass of faint SNe~IIn, we now add SN~2008S
in NGC~6946 ($d = 5.6$ Mpc; Sahu et al.\ 2006), discovered on 2008
Feb.\ 1.8 UT (Arbour \& Boles 2008).  It is of particular interest
because Prieto et al.\ (2008) found an associated infrared (IR) source
in pre-explosion {\it Spitzer} images.  IR data and visual upper
limits suggest that the progenitor was obscured by circumstellar dust
and had a modest mass of only 10--20 M$_{\odot}$ (Prieto et al.\
2008), below the range of initial masses usually attributed to LBVs
(Smith et al.\ 2004; Smith 2007).  Prieto et al.\ (2008) favor a mass
at the lower end of this range, with certain assumptions.  Following
the report of a similar obscured progenitor of a transient in NGC~300
(Prieto 2008), Thompson et al.\ (2008) proposed that these two objects
constitute a new class of transients, perhaps related to
electron-capture SNe in stars with initial mass $\sim$9 M$_{\odot}$.
Here we study the outburst of SN~2008S.\footnote{After preprints of
our paper appeared, Bond et al.\ (2009) and Berger et al.\ (2009)
reported extensive observations of a transient in NGC~300 that appears
to be very similar to SN~2008S.}  We find that despite the unusual
low-luminosity progenitor, the {\it outburst} resembles known SN
impostors, but is unlike any other objects that have been proposed as
weak core-collapse SNe or electron capture SNe.

\begin{deluxetable}{rccccc}
\tablewidth{0pc}\tighten
\tablecaption{Photometry of SN~2008S}
\tablehead{
\colhead{JD\tablenotemark{a}}& \colhead{$B$} & \colhead{$V$}& \colhead{$R$}&
\colhead{$I$} & Tel.\tablenotemark{b}
}
\startdata
  1.20 &   $-$     &   $-$     &  16.35(07) &   $-$     &M\\
  1.27 & 18.03(06) &   $-$     &    $-$     &   $-$     &U\\
  2.55 & 17.94(04) & 17.00(04) &    $-$     &   $-$     &U\\
  3.20 &   $-$     &   $-$     &  16.35(05) &   $-$     &M\\
  5.03 & 17.79(08) & 16.97(08) &    $-$     &   $-$     &U\\
  5.20 &   $-$     &   $-$     &  16.33(06) &   $-$     &M\\
  6.20 &   $-$     &   $-$     &  16.30(05) &   $-$     &M\\
  7.20 &   $-$     &   $-$     &  16.32(07) &   $-$     &M\\
  8.20 &   $-$     &   $-$     &  16.33(07) &   $-$     &M\\
  9.20 &   $-$     &   $-$     &  16.40(05) &   $-$     &M\\
 10.52 & 17.80(05) & 16.88(05) &    $-$     &   $-$     &U\\
 12.20 &   $-$     &   $-$     &  16.39(07) &   $-$     &M\\
 23.20 &   $-$     &   $-$     &  16.49(07) &   $-$     &M\\
 31.03 & 18.18(07) & 17.40(04) &  16.80(03) & 16.35(08) &N\\
 33.20 &   $-$     &   $-$     &  16.80(11) &   $-$     &M\\
 69.00 & 19.82(04) & 18.57(02) &  17.79(02) & 17.04(02) &N\\
 75.94 &   $-$     & 18.98(17) &  17.85(06) & 17.24(07) &N\\
 83.00 & 20.51(08) & 19.14(04) &  18.25(02) & 17.45(02) &N\\
 89.95 & 20.74(10) & 19.49(02) &  18.52(02) & 17.62(02) &N\\
 94.97 & 20.98(09) & 19.69(05) &  18.65(02) & 17.78(02) &N\\
 95.93 & 21.03(17) & 19.75(05) &  18.72(03) & 17.86(04) &N\\
117.96 & 21.73(16) & 20.66(13) &  19.58(08) & 18.58(05) &N\\
127.97 &   $-$     & 20.51(23) &  19.69(13) & 18.82(07) &N\\
135.94 & 21.83(23) & 21.20(23) &  19.94(10) & 18.85(11) &N\\
138.93 &   $-$     & 20.81(36) &  19.80(15) & 18.89(11) &N\\
142.93 & 22.21(18) & 21.06(18) &  19.96(09) & 18.94(05) &N\\
267.20 &   $-$     &   $-$     &  21.17(13) &   $-$     &K\\
\enddata
\tablecomments{1$\sigma$ uncertainties (in units of 0.01 mag) are in 
parentheses.  There are also $U$-band observations of SN~2008S
with UVOT: JD = 1.27, $U = 18.55(10)$; JD = 2.55, $U = 18.39(08)$; 
JD = 5.03, $U = 18.19(10)$; JD = 10.52, $U = 18.41(08)$ mag.}
\tablenotetext{a}{Julian Date $-$ 2,454,000; add 3 for days since discovery.}
\tablenotetext{b}{M: 0.35~m Celestron telescope owned by M.P.M.,
unfiltered and calibrated to the $R$ band; U: {\it Swift}/UVOT; 
N: Lick Observatory 1-m Nickel telescope; K: Keck-I unfiltered image.}
\end{deluxetable}

\begin{figure}
\epsscale{1.04}
\plotone{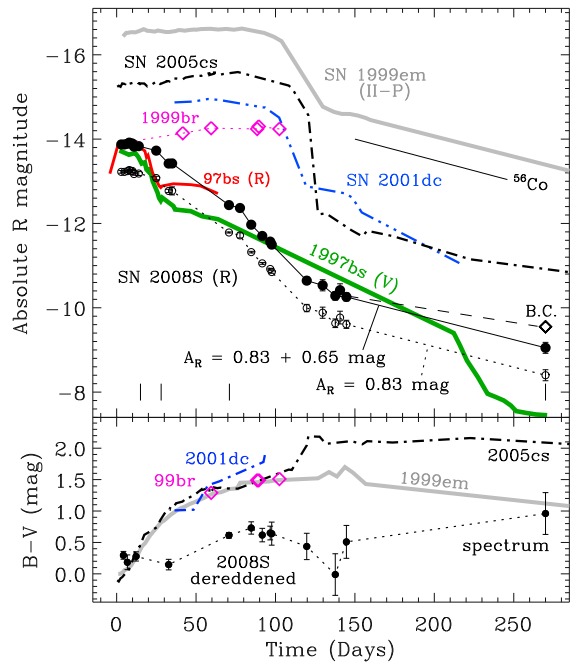}
\caption{{\it Top:} Absolute light curve of SN 2008S (adopting $d =
  5.6$ Mpc) corrected for Galactic extinction (open circles) of $A_R =
  0.83$ mag, and with an additional correction for local extinction
  (filled circles) of $A_R = 0.65$ mag (or $A_V = 0.87$ and $E(B-V) =
  0.28$ mag; see text).  For comparison, we show light curves for the
  normal SN~II-P 1999em (gray; Leonard et al.\ 2002), the
  low-luminosity SNe~II-P 1999br (magenta diamonds), 2001dc (blue;
  3-dot dash), and 2005cs (Pastorello et al.\ 2004, 2009), and the SN
  ``impostor'' SN~1997bs ($R$=thin red line; $V$=thick green line;
  from Van Dyk et al.\ 2000). Tick marks at the bottom show the epochs
  for which we have spectra.  The $^{56}$Co decay rate is also shown.
  The black diamond and dashed line correspond to the latest
  photometric point for SN~2008S with a bolometric correction (B.C.)
  applied for its very red color.  {\it Bottom:} The observed $B-V$
  color curve of SN~2008S corrected for $E(B-V)$=0.64 (Galactic +
  local), compared to SNe~II-P 1999em, 1999br, 2001dc, and 2005cs.}
\label{fig:one}
\end{figure}

\section{OBSERVATIONS}

We obtained photometry of SN~2008S with a 0.35~m Celestron telescope
(M.P.M.), the {\it Swift} Ultraviolet/Optical Telescope (UVOT; Roming
et al.\ 2005), and the Lick Observatory 1-m Nickel telescope. The
field of SN~2008S has been calibrated because its host galaxy produced
several SNe over the past decade. Unfiltered M.P.M.\ data are treated
as roughly $R$ band. For the M.P.M. and Nickel data, we used
point-spread-function (PSF) fitting to perform photometry.
For the UVOT reductions, we employed the photometric calibration
described by Li et al.\ (2006). Final photometry of SN~2008S is
reported in Table~1. Figure~1 shows the $R$-band light curve and $B-V$
color curve.

\begin{figure}
\epsscale{0.96}
\plotone{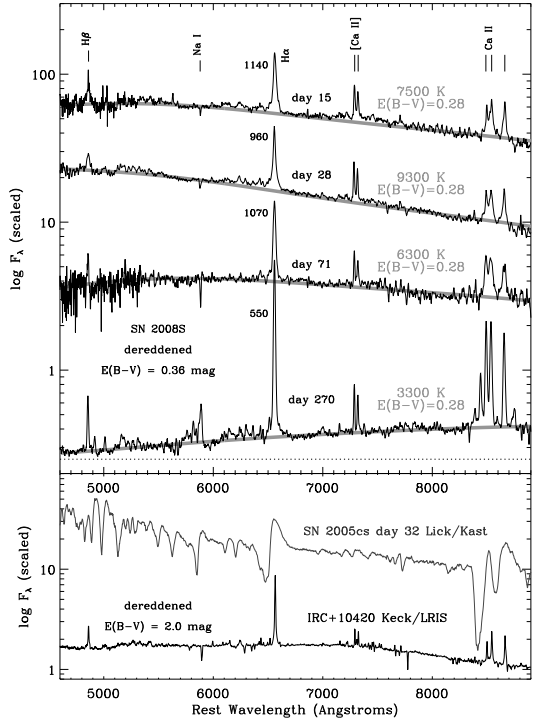}
\caption{{\it Top:} The spectra in black are of SN~2008S corrected for
  Galactic extinction only, assuming $E(B-V) = 0.36$ mag.  The days
  corresponding to each spectrum are after discovery.  The gray curves
  show blackbodies at the indicated temperatures reddened by
  $E(B-V) = 0.28$ mag to mimic additional local extinction.  The numbers
  left of the H$\alpha$ line indicate FWHM values in km s$^{-1}$,
  although note that these are probably overestimates for the first
  three epochs which have lower resolution. {\it Bottom:} An
  unpublished day 32 spectrum of the low-luminosity SN~II-P 2005cs
  (gray) from our spectral database, obtained at Lick Observatory, and
  a Keck/LRIS spectrum of IRC+10420 obtained on 2008 Oct.\ 28,
  dereddened by $E(B-V) = 2$ mag for comparison.}
\label{fig:two}
\end{figure}

We also obtained visual-wavelength spectra of SN~2008S using the Kast
double spectrograph (Miller \& Stone 1993) on the 3-m Shane reflector
at Lick Observatory (days 15, 28, and 71 after discovery), and using
the Low Resolution Imaging Spectrometer (LRIS; Oke et al.\ 1995) at
Keck Observatory (day 270). The slit of width $\sim $2\arcsec\ at Lick
and 1\arcsec\ at Keck yielded spectral resolving powers at red
wavelengths of $\lambda/\Delta \lambda \approx 670$ and 1250,
respectively, and it was oriented along the parallactic angle to
minimize the effects of atmospheric dispersion (Filippenko 1982).
Standard spectral data reduction was performed.  The spectra are
plotted in Figure~\ref{fig:two} after having been corrected for
Galactic reddening of $E(B-V) = 0.36$ mag (Schlegel, Finkbeiner, \&
Davis 1998).  We used the unfiltered guide-camera image from Keck LRIS
observations on day 270 to derive the last entry in Table 1, by
comparison with calibrated field stars.  The last $B-V$ measurement in
Figure~\ref{fig:one} is an approximate value derived from the
continuum shape in the day 270 spectrum, corrected for reddening (see
below).

Our spectra have insufficient dispersion to infer the local reddening
of SN 2008S using Na~{\sc i}~D absorption, and this would be
problematic anyway because the line changes to emission at late times.
In this paper, we draw attention to the remarkable similarity between
the spectrum of SN~2008S and that of the Galactic hypergiant star
IRC+10420; the basis for and implications of this comparison will be
discussed in more detail below (see Fig.~\ref{fig:two}).  For now, we
mention the spectroscopic similarity only as a basis for assuming that
the intrinsic values of $T_{\rm eff}$ for these two objects are
similar, in order to constrain the reddening.  The spectral type of
IRC+10420 is currently mid-A (Oudmaijer 1998).  Thus, with $T_{\rm
  eff} \approx 7500$~K, the observed continuum shape of SN~2008S near
peak implies a local extinction of $E(B-V) \approx 0.28$ mag in
addition to the Galactic extinction already applied in
Figure~\ref{fig:two} (this extinction or the intrinsic $T_{\rm eff}$
seems to vary as the object evolves). This local reddening is
consistent with the $B-V$ color near maximum light: an intrinsic $B-V$
color of $\sim$0.2 mag for 7500~K accounts for the corrected $B-V$ in
Fig.~\ref{fig:one}.

\section{BASIC PARAMETERS}

The extinction-corrected light curve in Figure~\ref{fig:one} shows a
peak $M_R = -13.9$ mag, corresponding to $L_{\rm peak} = 3 \times
10^7$ L$_{\odot}$ with zero bolometric correction.  For $T_{\rm eff} =
7500$~K, the emitting radius at peak is $2.3 \times 10^{14}$~cm or
15~AU.  Modest expansion speeds of (full width at half-maximum
intensity [FWHM] $\la$1000 km s$^{-1}$) near peak are indicated by
H$\alpha$ and Ca~{\sc ii} lines, while [Ca~{\sc ii}] lines are
narrower, appearing unresolved even in our higher-resolution LRIS
spectra with FWHM $\la$240 km s$^{-1}$. The broader line wings of
H$\alpha$ and Ca~{\sc ii} likely arise from multiple scattering by
thermal electrons in the dense wind (e.g., Chugai 2001).  Integrating
the light curve in Figure 1 and assuming zero bolometric correction,
the total radiated energy in the first 270 days is $E_{\rm rad} =
10^{47.8}$ ergs, between those of SN~1954J ($10^{47.3}$ ergs) and
P~Cyg's 1600 outburst ($10^{48.4}$ ergs) (Humphreys et al.\ 1999).

The peak of SN~2008S approaches the bottom end of the luminosity
distribution for core-collapse SNe~II.  This low end constitutes a
subset of low-luminosity SNe~II-P thought to have kinetic
energies $\ga$10$^{50}$ ergs, including objects such as SNe~1994N,
1999br, 1999eu, 2001gc, and 2005cs (see Pastorello et al.\ 2004).
Aside from the partial overlap in peak luminosity, however, SN~2008S
shares little else in common with this class.  Their light curves all
exhibit flat plateaus, ending with a sharp decline at $t = 100$--120
days to a $^{56}$Co decay tail, marking their transition from
photospheric to nebular phases.  SN~2008S, in contrast, shows a
relatively linear decline from peak with no clear transition of this
type. Moreover, the color evolution of SN~2008S is nothing like this
class of low-luminosity SNe~II-P (Fig.~\ref{fig:one}), whose color
evolution is like that of the more normal-luminosity SNe~II-P,
with little dispersion in the group (see Pastorello et al.\ 2004).
The late-time $R$-band decline of SN~2008S (Fig.~\ref{fig:one})
roughly matches the rate of 0.01 mag day$^{-1}$ that one expects for
$^{56}$Co decay from 0.002 M$_{\odot}$ of $^{56}$Ni.  However,
Figure~\ref{fig:one} shows only the $R$ magnitude with {\it no
bolometric correction}.  The day 270 spectrum has a very red continuum
peaking at $\ga$1~$\micron$ even after correcting for reddening, so
the substantial bolometric correction raises the luminosity by
$\sim$0.5 mag and makes the true decay rate about 0.06 mag day$^{-1}$
-- almost half that of $^{56}$Co.  The late-time decline rate of a SN
can be faster than the $^{56}$Co decay rate with energy leakage, but
it cannot be slower if radioactivity is the source.  It is therefore
unlikely that the late-time luminosity of SN~2008S is powered by
radioactive decay.

Furthermore, the spectral evolution of SN~2008S is unlike those of the
class of weak core-collapse SNe, all of which exhibit remakably
homogeneous spectral properties (Pastorello et al.\
2004).\footnote{Indeed, Pastorello et al.\ (2004) suggest that their
spectral homogeneity is sufficient to infer the phase in cases when the
explosion date is not known.} The observed ejecta speeds in faint
SNe~II are slower than those of normal SNe~II-P, typically declining
from 5000 to 3000 km s$^{-1}$ in the first months, and reaching
1000--1500 km s$^{-1}$ by day 100.  The speeds seen in SN~2008S were
even slower, however, dropping from $\sim$1000 km s$^{-1}$ at early
times to below 600 km s$^{-1}$ (Fig.~\ref{fig:two}).  The smooth
continuum and bright narrow emission lines of H and Ca~{\sc ii} at all
times in SN~2008S are quite unlike the strong absorption lines and
broad P~Cyg profiles observed in the photospheres of faint
core-collapse SNe (see SN~2005cs in Fig.~\ref{fig:two}).  Finally,
SN~2008S did not show the characteristic transition from photospheric
to nebular phases seen in the low-luminosity SNe~II; instead, its day
270 spectrum was almost identical to, although redder than, its
spectrum near peak.  Thus, we find that the spectrum of SN~2008S did
not arise from a recombination front receding through expanding and
cooling ejecta, as in weak core-collapse SNe~II.  Its spectrum points
to a different mechanism.\footnote{It is hard to definitively
eliminate the possibility that SN~2008S arose from a very weak
($<$10$^{49}$ erg) core-collapse SN shock interacting with dense
circumstellar material (CSM), as in more luminous SNe~IIn, but this
scenario would require the underlying SN photosphere to be at least a
factor of 100 fainter than a normal SN~II-P and would be
indistinguishable from CSM interaction caused by a non-terminal
explosion.}

Thompson et al.\ (2008) raised the question of whether SN~2008S and
the NGC~300 transient may have been electron-capture SNe (ecSNe) from
stars of initial mass $\sim$9 M$_{\odot}$.  While it is admittely
unclear exactly what the observed properties of ecSNe should be, one
might expect an ecSN from a 9~M$_{\odot}$ super-asymptotic-giant
branch star to resemble a weak core-collapse SN in a $\sim$10
M$_{\odot}$ red supergiant, since both have rapidly expanding H
envelopes with similar mass.  Indeed, some authors have proposed that
the class of weak SNe~II-P may be ecSNe (Chugai \& Utrobin 2000;
Hendry et al.\ 2005; Kitaura et al.\ 2006).  Since the observed
parameters of SN~2008S are quite distinct from those of the weak
SNe~II, as discussed above, we consider it unlikely that SN~2008S was
an ecSN.

On the other hand, the low peak luminosity of SN~2008S is consistent
with the observed range for SN impostors.  Despite differences in the
progenitor stars, the light curve of SN~2008S closely matches that of
the SN impostor SN~1997bs (Fig.~\ref{fig:one}), for which the
progenitor and surviving star were detected (Van Dyk et al.\ 2000; but
see Li et al.\ 2002).  Similarly, the slow expansion speeds, strong
Balmer lines, and smooth continuum match those observed in nearby LBVs
and other SN impostors, supporting our earlier conjecture that
SN~2008S was not a core-collapse SN (Steele et al.\ 2008).  The
Eddington parameter, $\Gamma =(\kappa_e L)/(4\pi GMc)$, is the factor
by which a star exceeds the classical Eddington limit, assuming that
Thomson scattering ($\kappa_e \approx 0.34$) dominates the opacity.
With such a high value of $L_{\rm peak} = 3 \times 10^7$ L$_{\odot}$,
SN~2008S would have $\Gamma \approx 40$ ($M$/20 M$_{\odot})^{-1}$.
This huge Eddington parameter is a factor of $\sim$10 higher than that
of $\eta$ Car during its 1843 eruption, when the star shed $\sim$10
M$_{\odot}$ in a few years (Smith et al.\ 2003).  This high Eddington
ratio may hint that the SN~2008S event was explosive, although that
does not necessarily imply that it was not an LBV-like event.  For
example, the same 1843 eruption of $\eta$ Car produced a fast blast
wave of $\sim$5,000 km s$^{-1}$ and $\sim$10$^{50}$ ergs (see Smith
2008).

\section{DISCUSSION}

We point out an uncanny similarity between the visual-wavelength
spectra of SN~2008S near peak luminosity and the spectrum of the
Galactic hypergiant star IRC+10420 (see Fig.~\ref{fig:two}).  Both
objects exhibit a smooth continuum dominated by narrow H$\alpha$,
[Ca~{\sc ii}], and Ca~{\sc ii} emission.  Such strong, narrow Ca~{\sc
ii} emission lines have not been seen before in a SN or SN impostor,
and are extremely rare among known stars. The strong Ca~{\sc ii} and
[Ca~{\sc ii}] lines have not been seen in SNe IIn where the radiation
is produced by shock interaction with a dense CSM.  IRC+10420 is an
evolved massive star in a yellow (spectral type of mid-A) hypergiant
phase with strong mass loss (Humphreys, Davidson, \& Smith 2002).
This phase may be a counterpart to the LBVs at cooler $T_{\rm eff}$
(Smith et al.\ 2004).

This spectral similarity between SN~2008S and IRC+10420 does not
necessarily mean that the objects are in the same evolutionary phase.
Although IRC+10420 has experienced strong variability in the past 30
years, it is not currently in a giant eruption.  Nevertheless, the
similarity does indicate similar values of $T_{\rm eff}$ for SN~2008S
in outburst and IRC+10420 in its current quiescent state.  It also
demonstrates that the observed spectrum of SN~2008S can plausibly
originate in an opaque and turbulent wind, because this is known to be
the case for IRC+10420, whereas no such precedent exists for
core-collapse SNe.  IRC+10420's mass-loss rate is estimated as $\sim
10^{-3}$ M$_{\odot}$ yr$^{-1}$ (see Humphreys et al.\ 2002).  The
Ca~{\sc ii} lines imply that both IRC+10420 and the progenitor of
SN~2008S are (or were) obscured by dust (Jones et al.\ 1993; Prieto et
al.\ 2008), since one potential explanation for the unusually strong
Ca~{\sc ii} lines is that radiation from SN~2008S might have vaporized
grains that were previously in equilibrium around a less luminous
progenitor.

Effective temperatures around 7500~K in both objects imply an
interesting regime where H is recombining, with serious consequences
in a wind driven by radiation force.  As such, the classical
electron-scattering Eddington limit may be altered in the outermost
layers of the star or inner wind.  If H recombines in the outflow, the
opacity will drop and the radiation field may no longer be able to
effectively impart momentum to the outflowing material.  An
inhomogeneous wind may stall or partly fail, and some material may
fall back onto the star, as has been suggested for IRC+10420
(Humphreys et al.\ 2002).

Recent numerical simulations of super-Eddington winds show a complex
pattern of outflow and infall (van Marle, Owocki, \& Shaviv 2008).
The general character of the winds in these simulations closely
matches the situation we envision for SN~2008S, although it is
difficult to evaluate this comparison quantitatively.  If our
suggested picture of a failed super-Eddington wind is applicable, we
might expect high-resolution spectra of the H$\alpha$ and Ca~{\sc ii}
lines to reveal signatures of simultaneous outflow and infall such as
inverse P~Cyg features, and asymmetric or double-peaked profiles
caused by self absorption.  These have in fact been seen in IRC+10420
(Oudmaijer 1998; Humphreys et al.\ 2002).  Following our initial
prediction, these types of line profiles were indeed shown in
higher-resolution spectra of the related transient in NGC~300 (Berger
et al.\ 2009; Bond et al.\ 2009).  Unfortunately, our low-dispersion
spectra do not resolve the detailed line-profile shapes in SN~2008S
itself.  A super-Eddington wind with $\Gamma > 10$ can drive strong
mass loss, with rates up to $\sim$0.1 M$_{\odot}$ yr$^{-1}$ (Owocki et
al.\ 2004).  If the ratio of radiated energy to kinetic energy is near
unity, as in $\eta$ Car (Smith et al.\ 2003; Smith 2006), then we
might expect an ejected mass of $\sim$0.16$\pm$0.05 $M_{\odot}$.


The LBV outburst phenomenon has been observed to occur in stars with
initial masses down to about 20 M$_{\odot}$ (e.g., Smith et al.\
2004).  For the lower mass range, the instability is thought to result
from heavy mass loss as a red supergiant (RSG), so that the stars have
high $L/M$ ratios in post-RSG blue loops (see Smith et al.\ 2004).  If
a similar mechanism operates in SN~2008S, as we suggest, then the
relatively low initial mass of 10--20 M$_{\odot}$ inferred for the
progenitor star by Prieto et al.\ (2008) has important
implications. It tells us that the phenomenon of episodic pre-SN mass
loss seen in LBVs can extend to even lower-mass stars than previously
thought, perhaps down to 15 M$_{\odot}$ or less.  At such low masses,
it is not at all clear that the star needs to be a blue supergiant to
experience a similar type of non-terminal outburst.  The mass loss
that caused the self obscuration may have occurred as a RSG with a
high $L/M$ value (Heger et al.\ 1997), possibly initiating the star's
consequent instability, whereas is seems likely that the high
luminosity of the event itself evaporated most of that dust.  The
expected short duration of $\lesssim 10^4$ yr for that preceding phase
(Heger et al.\ 1997) satisfies the expectations of Thompson et al.\
(2008) that these obscured phases would be short lived.  If SN~2008S
was a super-Eddington event in a $M_{\rm ZAMS} \la 20$~M$_{\odot}$
star, it strengthens the hypothesis that the progenitor of SN~1987A
ejected its nebula in a similar outburst (Smith 2007).

Thompson et al.\ (2008) proposed that SN~2008S and the similar event
in NGC~300 represent a new class of transients, while our observations
of the SN~2008S outburst itself imply the somewhat different
interpretation that they extend the parameter space of the already
diverse class of SN impostor outbursts to include progenitors with
lower luminosity than previously thought.  Since the underlying
mechanism that triggers LBV eruptions is still not diagnosed, it is
difficult to be certain about the difference in these perspectives.
We suspect that SN~2008S and the transient in NGC~300 are special
cases where the progenitors were highly obscured because of recent
mass loss and because of their relatively low progenitor mass. The
scenario of a non-terminal explosion or ``SN impostor'' predicts that
the star will be detectable again, although it may take decades before
the star recovers to thermal equilibrium.

\acknowledgments 
\scriptsize

N.S.\ acknowledges interesting discussions concerning the nature of SN
2008S with J.\ Prieto and T.\ Thompson, and of super-Eddington winds
with S. Owocki. Some of the data presented herein were obtained at the
W. M.\ Keck Observatory, which is operated as a scientific partnership
among the California Institute of Technology, the University of
California, and NASA; the observatory was made possible by the
generous financial support of the W. M.\ Keck Foundation.  We thank
the Lick and Keck Observatory staffs for their dedicated help, as well
as the following for their assistance with some of the observations:
C.\ Anderson, A.\ Barth, J.\ Chu, J.\ Leja, B.\ Macomber, B.\ Tucker,
and J.\ Walsh.  N.S.\ was partially supported by NASA through grants
GO-10241 and GO-10475 from the Space Telescope Science Institute,
which is operated by AURA, Inc., under NASA contract NAS5-26555, and
through Spitzer grants 1264318 amd 30348 administered by JPL.
A.V.F.'s supernova group at U.C.\ Berkeley is supported by NSF grant
AST-0607485, the TABASGO Foundation, and the Richard \& Rhoda Goldman
fund. We acknowledge the use of public data from the {\it Swift}
archive.

{\it Facilities:} Keck I (LRIS), Lick 3-m (Kast), Lick 1-m, Lick KAIT,
{\it Swift} UVOT.


\end{document}